\documentclass[apj]{emulateapj}

\usepackage{graphicx}

\shorttitle{Comparing the energy spectra of UHECRs from EAS arrays}
\shortauthors{Ivanov}

\begin{document}

\title{Comparing the energy spectra of ultra-high energy cosmic rays\\ measured with EAS arrays}

\author{A.A. Ivanov}
\affil{Shafer Institute for Cosmophysical Research and Aeronomy,
31 Lenin Avenue, Yakutsk 677980, Russia}
\email{ivanov@ikfia.ysn.ru}

\begin{abstract}
The energy spectra of ultra-high energy cosmic rays (CRs) measured with giant extensive air shower (EAS) arrays exhibit discrepancies between the flux intensities and/or estimated CR energies exceeding experimental errors. The well-known intensity correction factor due to the dispersion of the measured quantity in the presence of a rapidly falling energy spectrum is insufficient to explain the divergence. Another source of systematic energy determination error is proposed concerning the charged particle density measured with the surface arrays, which arises due to simplifications
(namely, the superposition approximation)
 in nucleus-nucleus interaction description applied to the shower modeling. Making use of the essential correction factors results in congruous CR energy spectra within experimental errors. Residual differences in the energy scales of giant arrays can be attributed to the actual overall accuracy of the EAS detection technique used. CR acceleration and propagation model simulations using the dip and ankle scenarios of the transition from galactic to extragalactic CR components are in agreement with the combined energy spectrum observed with EAS arrays.
\end{abstract}


\section{Introduction}
Ultra-high energy cosmic rays (UHECRs) are presently measured using a number of giant extensive air shower (EAS) arrays. The observed energy spectrum exhibits the cutoff predicted by \citet{G} and \citet{ZK} (GZK) and `Ankle' features at energies of approximately $4\times10^{19}$ eV and $5\times10^{18}$ eV \citep{Fksm}. However, there are essential discrepancies between the flux intensities and/or the estimated energies of the initial cosmic rays (CRs) generating EASs detected by different arrays. These discrepancies exceed instrumental errors and
it makes it difficult to decide for
the validity of the results obtained.

The current paper presents an analysis of the sources of these discrepancies. It is shown by i) correcting the CR intensity due to instrumental errors and power law spectrum and ii) taking into account model uncertainty in the estimation of EAS initial nucleus energy that the observed UHECR energy spectra appear to be congruent. Residual differences in UHECR energy scales of arrays can be attributed to the actual overall accuracy of the EAS detection technique.

\section{Techniques used to measure the energy spectrum of UHECRs}
There are two basic techniques for measuring UHECR parameters with EAS arrays and, in particular, for estimating the energy. The first is the measurement of the electromagnetic component and/or muons reaching the ground; the energy is estimated using a model simulation of the particle density at a particular distance from the shower axis (e.g., $S_{600}$). The second technique is based on the measurement of the ionization integral of the longitudinal EAS profile with fluorescence or Cherenkov light detectors. In this case the UHECR energy, $E_0$, is estimated as a sum of the ionization integral, $E_i$, and the `missing energy', $E_m$, carried by hadrons, muons, and neutrinos. The ionization integral is given by
$$
E_i=\frac{\epsilon}{t_0}\int N_e(t)dt,
$$
where $N_e(t)$ is the number of electrons and positrons at depth $t$; $\epsilon$ is the critical energy in air; $t_0$ is the electron radiation length in air. The missing energy is comparatively small ($E_m/E_0<0.1$) at energies above 1 EeV (=$10^{18}$ eV), so the method used can be considered to be nearly model-independent.

A typical example of an instrument applying the first technique is the Akeno Giant Air Shower Array (AGASA) (e.g., \cite{AGASASpectrum}), while the second approach is realized in the High Resolution Fly's Eye (HiRes) array, consisting of fluorescence light detectors (\cite{HiRes08} and references therein). Next-generation arrays combine both techniques: The Pierre Auger Observatory (PAO;~\citet{PAO09}) and the Telescope Array (TA;~\citet{TA09}) comprise charged particle detectors and fluorescence telescopes.

In the Yakutsk array experiment these two techniques are also realized: There are scintillators on the ground detecting electrons, positrons, photons, and muons; scintillators beneath the ground detecting muons; and photomultiplier tubes (PMTs) detecting the air Cherenkov light produced by the showers \citep{CRIS}. Data from the scintillators and PMTs are used to estimate the energy of the UHECR particles initiating EASs.
It has been shown that the two independent estimates diverge significantly,
the shift in the $\lg E$ between resultant spectra of CRs
is approximately 0.12 in the range $E_0>10^{18}$ eV~\citep{NJP}.

\begin{figure}[t]
\includegraphics[width=\columnwidth]{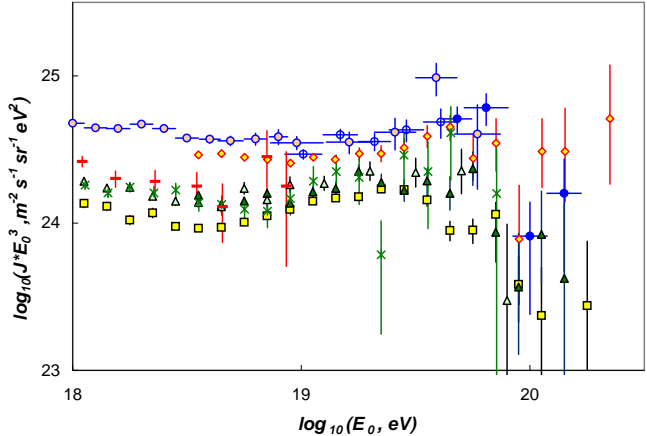}
\caption{The differential UHECR flux (multiplied by $E_0^3$ and correction factors from Table~\ref{Table:RJ}) measured with EAS arrays. Experimental data are from AGASA (rhombuses, \citet{AGASASpectrum}), HiRes I, II (open and filled triangles, \citet{HiRes08}), TA (crosses, \citet{TA09}), Haverah Park (HP, horizontal bars, \citet{HP}), PAO (squares, \citet{PAO09}), the Yakutsk array (open, crossed and filled circles, \citet{CRIS}).}
\label{fig:SpectraRJ}
\end{figure}

\section{A correction to the CR intensity measured with EAS arrays}
A first step in comparing the observed energy spectra is to include a correction to the measured intensity of CRs caused by the instrumental errors and power spectrum.

It was shown by \citet{Ztspn} that there should be a difference between the observed intensity of CRs and the original intensity in the case of a rapidly falling energy spectrum, due to instrumental errors and fluctuations in the shower parameters measured.

Then \citet{Mrzn} and \citet{NNK} calculated the measured intensity in the case of a lognormal distribution of $S_{600}$ and the so-called shower size, $N_e$:
\begin{equation}
J(N_e)=J_0(N_e)\exp(\frac{\sigma_N^2\kappa(\kappa-a_N)}{2a_N^2}),
\label{Eq:NNK}
\end{equation}
where $J_0$ is the actual intensity; $\sigma_N$ is the RMS deviation of $\ln N_e$; $\kappa$ is the integral energy spectrum index; and $a_N=d\ln N_e/d\ln E_0$.

In our case, the target values are the parameter $\hat{E}$ = `primary particle energy' estimated after shower detection, and the actual energy of
the
CR, $E_0$, that initiated the EAS. The estimated energy has a distribution around the mean value formed by the instrumental errors and fluctuations with a RMS deviation, $\sigma$. The energy fluctuation is small in comparison with instrumental errors and can therefore be neglected.

If we assume the lognormal distribution of $y=\ln\hat{E}$, with an average value equal to $\ln E_0$, then the observed intensity of CRs is given by the convolution of the primary spectrum, $J_0\exp(-\kappa z)$, and the distribution of instrumental errors:
\begin{equation}
\hat{J}(z)=J_0\int_{-\infty}^\infty \exp(-\kappa z+\kappa y)\frac{\exp(-\frac{y^2}{2\sigma^2})}{\sqrt{2\pi}\sigma}dy.
\end{equation}
The resultant observed-to-initial intensity conversion factor is~\citep{Mrzn,NJP}:
$$R_J=J_0(z)/\hat{J}(z)=\exp(-\frac{\sigma^2\kappa^2}{2}).$$
The necessary conditions are a constant index and RMS error. As a crude approach, one can use the broken power law approximation of the energy spectrum given by the HiRes collaboration~\citep{HiRes08}, and the constant RMS error averaged in the range $E_0>10^{18}$ eV. In the vicinity
(with width $\sigma$)
of the break points the interpolation of the index can be used to prevent gaps in the spectrum. The procedure is inevitably iterative: The revised spectrum indexes (2.24, 1.8, and
4.5
below the ankle, between the ankle and the cutoff, and above the cutoff energies, respectively) rather than observed indexes should be used in a correction factor.

In Fig.~\ref{fig:SpectraRJ} the energy spectra presented were observed by giant EAS arrays with CR intensity correction factors calculated using the HiRes power law approximation of the energy spectrum and instrumental errors inherent to arrays (Table~\ref{Table:RJ}). In the PAO and HP cases, however, the corrections are already applied in the original works~\citep{HP,PAO09} so no intensity corrections are made.

\begin{table}[t]
\caption{
Instrumental errors, $\sigma$, and intensity correction
factors, $R_J$, for EAS arrays.
\label{Table:RJ}}
\begin{tabular}{lcccc}&&&&\\
\hline
                     Array & AGASA & HiRes &  TA  &  Yakutsk \\
\hline
              $\sigma$, \% &    25 &    23 &   23 &    32   \\
       $R_J(E_0<4.47$ EeV) &  0.86 &  0.88 & 0.88 &  0.77   \\
  $R_J(4.47<E_0<56.2$ EeV) &  0.90 &  0.92 & 0.92 &  0.85   \\
       $R_J(E_0>56.2$ EeV) &  0.53 &  0.59 & 0.59 &  0.36   \\
\hline
\end{tabular}
\end{table}

Concerning the HiRes experiment, the monocular reconstruction results~\citep{HiRes08} are used here to yield spectra with the best statistical power over a wide energy range. Energy estimation errors of the two HiRes detectors in the monocular mode are derived based on the original data \citep{HiRes09a}, with the ratio distribution of energies measured by HR1 and HR2 independently for the same EAS event. The RMS deviation of the ratio is found to be $0.33\pm 0.01$. An immediate consequence is the average energy estimation accuracy of the Fly's Eye detectors in the monocular mode, $\delta \hat{E}/\hat{E}\simeq \sqrt{0.5(\delta_{HR1}^2+\delta_{HR2}^2)}=0.23\pm 0.01$.

Since the preliminary results from the TA experiment~\citep{TA09} are obtained by fluorescence detectors in monocular mode applying the same data-handling procedure as in the HiRes case, the same energy estimation errors are assigned here to both arrays. For other experiments, instrumental errors in energy estimation are taken from the original papers.

\section{Energy scale difference due to the EAS modeling uncertainty}
The energy of the primary CR particle initiating the EAS is estimated using model relations between $E_0$ and measured shower parameters, such as the ionization integral, $E_i$, the particle density at 600 m from the axis, $S_{600}$, and the number of electrons at observational level, $N_e$. These relations are more or less dependent on the high-energy hadron--nucleus and nucleus--nucleus interaction models used. Thus, in addition to including instrumental errors, we must introduce a systematic `EAS modeling uncertainty' into the energy estimation procedure.

Existing inconsistencies in the interpretation of the experimental data on the primary energy and mass composition of UHECRs indicate the presence of this uncertainty. A number of attempts have been made to estimate the value of the effect and elucidate the source(s); a typical example is an analysis conducted by \citet{Knpp} of EAS simulations at energies above $10$ EeV.

\citet{Knpp} found a clear trend of convergence between different hadronic interaction models: The ionization integral differs by only 2\% in QGSJET 01, SIBYLL 2.1, and DPMJET 2.5, but the lateral distribution of EAS particles far from the core ($r>600$ m) still differs by about 15\% (photon and electron densities) and 30\% (muons) in the models. This means that there is an uncertainty in the energy determination at the surface water-Cherenkov detectors (HP, PAO) of about 20\%, and at the scintillation detectors (AGASA, Yakutsk) of $\sim15$\% for vertical showers and up to $30$\% for inclined showers where muons dominate the charged particle density at zenith angles $\theta>60^0$.

This estimate is undoubtedly a lower limit of the actual uncertainty, because it does not include the contribution from low-energy hadronic interactions and computational and coding errors. For instance, \citet{Knpp} have compared the performance of CORSIKA and AIRES codes for the same interaction model (QGSJET). The resultant mean positions of EAS maximum in atmosphere differ by about 25 g cm$^{-2}$, the electron numbers at the maximum differ by 6\%, while the photon and electron densities at $10^3$ m from the core differ by $\sim20$\%.

All these sources of energy estimation errors are symmetric in the sense that they can equiprobably increase or decrease the derived primary energy due to an inadequate interaction model or coding errors.

\begin{figure}[t]
\includegraphics[width=\columnwidth]{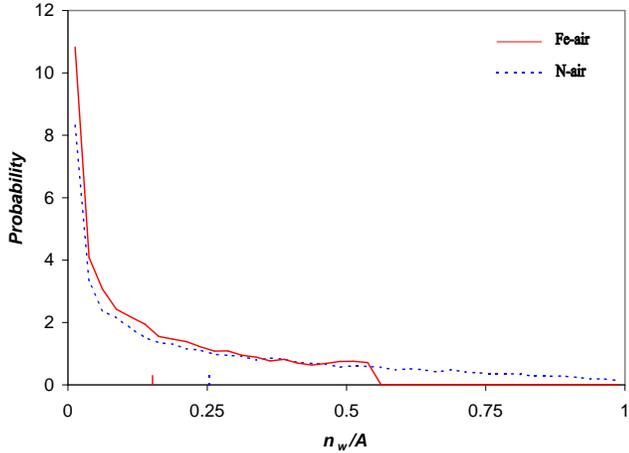}
\caption{Probability distribution of the wounded
(interacting)
nucleons fraction, $n_w/A$, in eikonal approximation. Calculation results for collisions of nitrogen and iron nuclei in air are shown. Average values are indicated by the vertical bars on the bottom scale.}
\label{fig:Wounded}
\end{figure}
However, there are other examples of the energy divergence sources that result in the systematic errors of $\hat{E}$. The first source is the
variation in the
mass composition of UHECRs. In this case we can use CORSIKA simulation results for the primary proton and iron nucleus (leaving aside more exotic primaries such as photons, neutrinos, etc.), QGSJET and SIBYLL models, applied to AGASA, Yakutsk array scintillation detectors~\citep{NgnQGS,Ddnk} and PAO fluorescence-detectors~\citep{Brbs}. In the case of iron primaries the energy estimate should be decreased by $\sim10$\% (AGASA), $\sim15$\% (Yakutsk) and $\sim3$\% (PAO) in comparison with proton-initiated showers. A corresponding correction in the HP case is applied by \citet{HP}.

Another example is the use of a common superposition model when treating nucleus-nucleus collisions in EAS simulation codes. While in some models (e.g. QGSJET, DPMJET) the variants of the nuclei fragmentation are implemented, in others (SIBYLL, HDPM) only the superposition approximation is used~\citep{Heck98}. In contrast, real nucleus--nucleus interactions are mostly peripheral, and the probability that the `wounded' (interacting) nucleon number, $n_w$, is equal to the projectile mass number $A$
(as in superposition approximation)
, is small. Fragmenting primary nucleus induces a shower of secondaries slowed down in comparison with EAS in the superposition model. This should result in $S_{600}^f>S_{600}^s$ and then a reduced primary energy estimate in the fragmentation model in comparison with the superposition model.

In the next section, EAS observables are estimated in a wounded nucleon model of nuclei interactions in air~\citep{WNM}.

\begin{figure}[t]
\includegraphics[width=\columnwidth]{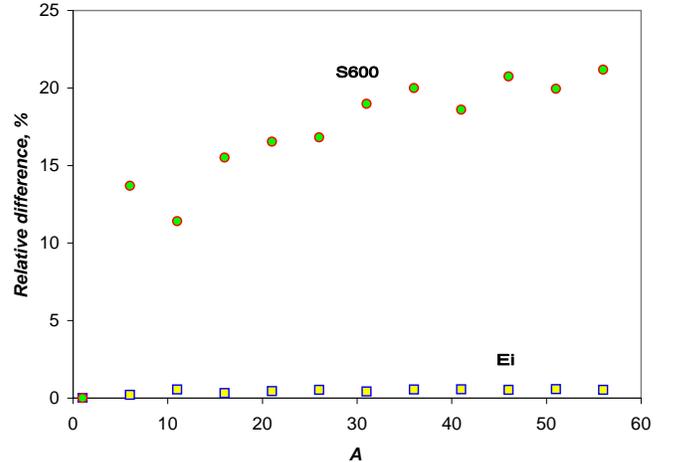}
\caption{Relative difference in charged-particle density, $S_{600}$ (circles), and ionization integral, $E_i$ (squares), in fragmentation and superposition models.}
\label{fig:Ratio}
\end{figure}

\section{Comparison of fragmentation and superposition models}
In the current work,
a simple eikonal approximation is used, implementing the Glauber model of nucleus--nucleus collisions. The geometric quantities--- average impact parameter, number of wounded/spectator nucleons, etc.---are calculated using an algorithm described by the PHOBOS collaboration~\citep{Alver08}. Only projectile nucleus fragmentation is considered, because the fragments of the target nuclei make an insignificant contribution to the cascade in the ultra-high energy domain. In Fig.~\ref{fig:Wounded} the distributions of wounded-nucleon number as a ratio to the mass number are given for Fe and N projectile nuclei.
Validity of the approximation used is based on the correct description of D-Au collisions at 200 GeV
in Relativistic Heavy Ion Collider~\citep{PHOBOS}.

Inelastic cross sections for projectile nuclei with $A\in(1,56)$ were calculated by the Glauber method \citep{Heck98}. The next step is to calculate the aggregate shower by adding the sub-cascades induced by the primary nucleus fragments at different levels in the atmosphere. A simplification here is that the spectator nucleons of the projectile are assumed to subsequently interact as a nucleus with the mass number $A-n_w$, while the wounded nucleons interact independently. For example, in the case of the primary nitrogen nucleus, we have a chain of N-B-Li-He-D-H nuclei interactions in air.

In order to derive $E_i$ and $S_{600}$ in EAS induced by fragmenting nuclei, we have to use the values as a function of the initial nucleon energy and sub-shower initial points, $x_i$. In the case of the ionization integral, calculations for three essentially different models have demonstrated~\citep{DAN} a small variation of $E_i$ ($\delta E_i/E_0<0.007$) in the energy interval ($0.1<E_0<10$) EeV and $x_i\in(0,200)$ g cm$^{-2}$.

Comparing the ionization integral in the fragmentation model
$$E_i^f=\sum_k E_i^k(\frac{n_w^kE_k}{A_k},x_i^k),$$
and superposition model
$$E_i^s=A E_i(\frac{E_0}{A},x_i^1),$$
we find a difference $(E_i^f-E_i^s)/E_0$ in the two models of less than 0.5\% for the mass number of primary nuclei in the interval (1,56).

In contrast, there is a considerable difference in the charged-particle densities of the two models. While $S_{600}$ is almost proportional to the energy of primary proton/iron nuclei according to the CORSIKA/SIBYLL and CORSIKA/QGSJET simulations  of \cite{NgnQGS} ($dlgE_0/dlgS_{600}^{\theta=0}=1.015\pm0.015$), the density rises with $x_i$, at least for the arrays at sea level. \cite{Cpdvll}\footnote{we have used here the attenuation length in $\sec\theta$ as an estimator of $dx_i/d\ln S_{600}$} showed that $\lambda_{S600}\sim400$ g cm$^{-2}$ for models of high multiplicity, and $\lambda_{S600}\sim300$ g cm$^{-2}$ for models of low multiplicity with the primary energy $E_0=1$ EeV.

Calculation of $S_{600}$ in the fragmentation and superposition models results in a $S_{600}^f/S_{600}^s$ ratio within the (1,1.17) interval for models of high multiplicity, and in (1,1.21) for models of low multiplicity. The relative difference $S_{600}^f/S_{600}^s-1$ in the latter case, and $(E_i^f-E_i^s)/E_0$, are shown in Fig.~\ref{fig:Ratio} as a function of the mass number of an EAS primary nucleus.

Using a superposition model to derive the energy of the primary nucleus from the measured shower parameters $E_i$ or $S_{600}$, we find that $\hat{E}$ is correct with accuracy better than $0.5$\% for $E_i$ measurement,
and is overestimated by the factor $R_S\in(1,1.2)$ in the case of $S_{600}$. The actual values depend on the mass composition of UHECR flux.

The measurements of the mass composition have been made by the HiRes collaboration in several
energy bins above 0.1 EeV: i) HiRes prototype and the MIA muon array data \citep{bZyyd} led to
the conclusion that CR intensity is changing from heavier to a lighter composition between
0.1 EeV and 1 EeV; ii) then, $X_{max}$ distribution width and elongation rate measurements
in the interval (1,2.5) EeV \citep{bbs} are found to be consistent with a constant or slowly
changing and predominantly light composition; iii) and finally, the recent measurements at
energies above 1.6 EeV \citep{HiRes09b} have found a proton dominated UHECR flux, within
QGSJET01 and QGSJET-II two-component proton-iron model assumptions.

On the contrary, PAO collaboration measurements of both the mean depth of shower maximum vs
energy and RMS for the $X_{max}$ distribution suggest the composition becoming heavier above
3 EeV \citep{Crnn}. In addition, the muon data recorded by the Yakutsk array indicate the
presence of a significant fraction of heavy nuclei in the CR flux at energies $E>10$ EeV
\citep{Glshkv}: The proton fraction is estimated as $0.32\leq f_P\leq 0.71$ (95\%CL) and iron fraction is $0.29\leq f_{Fe}\leq 0.68$ (95\%CL) in the two-component (P+Fe) composition and EPOS interaction model.

\begin{figure}[t]
\includegraphics[width=\columnwidth]{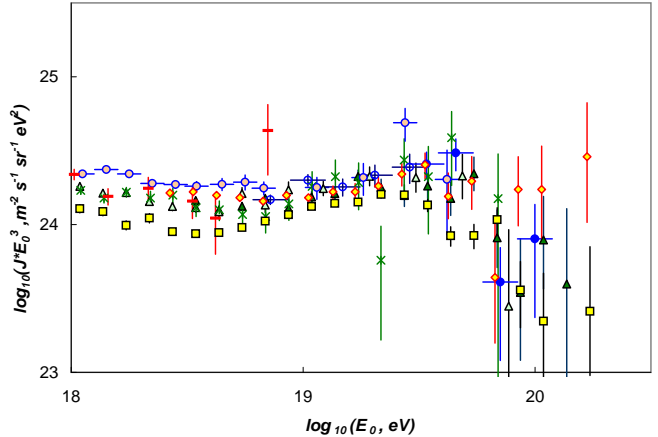}
\caption{The differential energy spectra measured with EAS arrays. Energy corrections are applied assuming EAS induced by a primary iron nucleus. CR intensities and data symbols are the same as in Fig.~\ref{fig:SpectraRJ}.}
\label{fig:SpectraRA}
\end{figure}

In this context we have to bear in mind both the possibilities of light and heavy UHECR compositions.
To illustrate the convergence/divergence of the energy spectra measured with EAS arrays, the
surface array energies are decreased\footnote{except HP data re-evaluated using CORSIKA/QGSJET~\citep{HP}} by the factor $1.2$ (Fig.~\ref{fig:SpectraRA}) as a lower limit to the primary energy in the case of the iron nucleus assumed as a primary; other case of the proton primary is illustrated in Fig.~\ref{fig:SpectraRJ}.
In other words, two figures (Figs. \ref{fig:SpectraRJ} and \ref{fig:SpectraRA})
represent the two extreme cases of UHECR nuclear compositions.

\section{Results and discussion}
It has been noted previously that UHECR energy spectra measured with giant EAS arrays agree with each other if the energy scales are adjusted (\cite{BW,DAN,Brznsk}, and other papers cited therein). However, the energy correction factors needed to merge the spectra exceed experimental errors. For example, \citet{Brznsk} used values 1.2, 0.75, 0.83, and 0.625 to shift PAO, AGASA, Akeno, and Yakutsk data to those of HiRes. Moreover, instrumental and modeling errors are believed to equiprobably increase or decrease the estimated energy. Instead, as \citet{Wtsn} concluded when comparing integral calibrated fluxes of UHECRs from giant arrays, $S_{600}$ measurements and fluorescence measurements of EAS assemble in two separate groups of data (Volcano Ranch, Haverah Park, AGASA vs PAO, HiRes results).

\begin{figure}[t]
\includegraphics[width=\columnwidth]{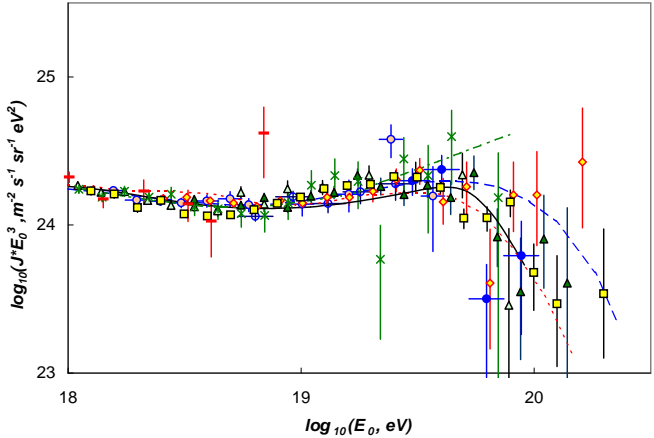}
\caption{The observed spectra with energy scales adjusted. Correction factors are taken from Table~\ref{Table:RA}. CR intensities and data symbols are the same as in Fig.~\ref{fig:SpectraRJ} and ~\ref{fig:SpectraRA}. Model calculation results: full line \citep{Dip}; dot and dash lines \citep{AGN}; dash-dot line \citep{Wbg}.}
\label{fig:SpectraRE}
\end{figure}

Considerations in this paper insist that, besides the `Moscow correction factor'\footnote{derived by \cite{Ztspn}, \cite{Mrzn}, and \cite{NNK}}, which reduces all CR intensities measured with EAS arrays, there is another factor that reduces the estimated energy of primary nuclei diminishing a difference between two groups of data (Fig.~\ref{fig:SpectraRA}). The remaining spread of $\hat{E}$ can be attributed to the real accuracy of the EAS measurement technique: instrumental errors + model uncertainty.

Two variants of the energy correction factors for EAS arrays are given in Table \ref{Table:RA} for the two extreme cases of the primary nuclei (H,Fe). Resultant spectra with adjusted energy scales are shown in Fig.~\ref{fig:SpectraRE}. Energy correction factors are assumed constant at energies above $1$ EeV, and the spectra are shifted to a sample mean.

Having an
indefinite
mass of EAS primary nuclei and model uncertainties to assign the energy correction factors, we can find an interval of values where $R_E$ would be. A sample of six experiments (Figs.~\ref{fig:SpectraRJ}, \ref{fig:SpectraRA}) estimates that the average energy determination error, $\delta \hat{E}/\hat{E}$, inherent to giant EAS arrays, is within an
(8,19)\%
interval.

Now we can compare the measured energy spectra with model simulations of extragalactic (EG) CRs predominating over the Galactic (G) component. The dip (described as `bump' in early measurements) observed in the CR spectrum~\citep{Bump1,Bump2}
was explained by \citet{Dip0}\footnote{it has been updated a number of times since; the latest is made by \citet{Dip}},
who studied the Bethe--Heitler pair production of protons from distant sources on the cosmic microwave background.

The authors conclude that the dip shape is formed by the universal modification factor and is independent of UHECR propagation details. The only phenomenon known to modify the dip is the presence of a significant fraction of nuclei in the primary beam.

In Fig.~\ref{fig:SpectraRE} the spectrum calculated in the dip scenario \citep{Dip} with protons accelerated in EG sources\footnote{acceleration spectrum index $\gamma=2.7$} is shown. Only the source luminosity is fitted to the experimental data. The position on the energy scale and shape of the dip agree with observed spectra within experimental errors, as well as the GZK effect. The excess-over-GZK flux observed by the AGASA array is considered to be the result of the primary-energy overestimation in inclined showers of the highest energies \citep{Cpdvll}.

\begin{table}[t]\begin{center}
\caption{Energy correction factors for EAS arrays.
\label{Table:RA}}
\begin{tabular}{lccccc}&&&&&\\
\hline
      Array & AGASA &  HP  & Yakutsk &  PAO & HiRes/TA\\
\hline
$R_E^{Fe}$ &  0.96 & 0.98 &  0.88  & 1.15 &  1.01\\
 $R_E^{H}$ &  0.81 & 1.01 &  0.71  & 1.29 &  1.09\\
\hline
\end{tabular}\end{center}
\end{table}

\citet{Wbg} have argued that the ankle in the spectrum marks the transition from G to EG components of CRs. This phenomenological `ankle scenario' uses a sum of power law G and EG spectra with slopes differing by $\Delta\gamma\sim1.8$. The result is shown in Fig.~\ref{fig:SpectraRE} by the dash-dot line. The authors emphasize the sharpness of the ankle observed, $d^2\ln JE^3/d\ln^2E$, which is consistent with the ankle scenario, while in the dip scenario the sharpness is of insufficient magnitude, especially in the case of a large nuclei-to-proton ratio in the primary beam.

The energy spectrum of UHECRs produced at the shock created by the expanding cocoons around active galactic nuclei combined with the G component of CRs produced in supernova remnants was calculated by \citet{AGN}. Expected CR composition shows an increase of $\overline{A}$ at $E_0\sim0.1$ EeV owing to the G component, and a second one at energy $\sim10$ EeV, produced by nonrelativistic cocoon shocks.

Calculated UHECR intensity \citep{AGN} as a function of energy is shown in Fig.~\ref{fig:SpectraRE} for the two cases: dip (dots) and ankle (dashed line) scenarios of the G to EG transition. Experimental errors are too large to distinguish between alternative scenarios.

\section{Summary}
A comparison is drawn between the energy spectra of UHECRs measured with giant EAS arrays, taking into account the necessary corrections to the measured CR intensity and the estimated energy of the primary particle. The intensity correction factor for the measured CR flux due to a rapidly falling energy spectrum and instrumental errors was derived in Moscow many years ago. EAS modeling uncertainty in the primary energy estimation is due to unknown nucleus--nucleus interaction characteristics at energies far beyond those studied in accelerator experiments.

In particular, there may be a difference between energies estimated in the superposition and fragmentation models of nucleus-nucleus interactions. It is shown that, indeed, using the superposition model to analyze the surface array data, e.g. $S_{600}$, we obtain $\hat{E}$, which has to be corrected due to the fragmentation rate overestimated.

Applying the essential correction factors, it is shown that UHECR energy spectra measured with EAS arrays are congruous within experimental errors arising from instrumental and model uncertainties. Residual differences in the energy scales of giant arrays are used to estimate UHECR energy determination error inherent in EAS detection techniques.

Model calculations in dip and ankle scenarios of the transition from G to EG components of CRs are in agreement with observed ankle and GZK features of the energy spectrum. However, the experimental uncertainties are too large to be able to distinguish between the scenarios. More data are needed from future arrays (Auger-North, satellite projects, etc.) to elucidate the details of the spectrum measured.

\acknowledgments
The author is grateful to the Yakutsk array staff for the data analysis and valuable discussions.
This work is supported in part by RFBR grant no. 09-02-12028 and by the Russian Federal Program `Scientific and Educational personnel' (contract no. 02.740.11.0248).

\end{document}